\documentclass[prd,amsmath,amssymb,superscriptaddress,floatfix,nofootinbib,10pt]{revtex4}
\usepackage{times}
\usepackage{amssymb,amsbsy,amsmath,amsfonts}
\usepackage{graphicx}
\usepackage{float}
\usepackage{color}
\usepackage{morefloats}
\usepackage{rotating}
\usepackage{srcltx}
\usepackage{slashed}
\usepackage{multirow}
\usepackage{verbatim}
\usepackage{hyperref}
\usepackage{tabularx}
\usepackage{bm}
\allowdisplaybreaks[4]

\usepackage{bbding}
\usepackage{threeparttable}

\DeclareUnicodeCharacter{2212}{\ensuremath{-}}

\newcommand{\PreserveBackslash}[1]{\let\temp=\\#1\let\\=\temp}
\newcolumntype{C}[1]{>{\PreserveBackslash\centering}p{#1}}
\newcolumntype{R}[1]{>{\PreserveBackslash\raggedleft}p{#1}}
\newcolumntype{L}[1]{>{\PreserveBackslash\raggedright}p{#1}}

\begin{document}

\title{ Determination of the $K^+\bar{K}^0$ scattering length and effective range from the  $D^+\to\bar{K}^0\pi^+\eta$ reaction}

\author{Jing Song}
\affiliation{School of Physics, Beihang University, Beijing, 102206, China}
\affiliation{Departamento de Física Teórica and IFIC, Centro Mixto Universidad de Valencia-CSIC Institutos de Investigación de Paterna, 46071 Valencia, Spain}

\author{Wei-Hong Liang}
\email[]{liangwh@gxnu.edu.cn}
\affiliation{Department of Physics, Guangxi Normal University, Guilin 541004, China}
\affiliation{Guangxi Key Laboratory of Nuclear Physics and Technology,
Guangxi Normal University, Guilin 541004, China}

\author{ Eulogio Oset}
\email[]{oset@ific.uv.es}
\affiliation{Departamento de Física Teórica and IFIC, Centro Mixto Universidad de Valencia-CSIC Institutos de Investigación de Paterna, 46071 Valencia, Spain}
\affiliation{Department of Physics, Guangxi Normal University, Guilin 541004, China}

\begin{abstract}
We study the scattering parameters of the \(K^+\bar{K}^0\) system through the analysis of the \(D^+\to\bar{K}^0\pi^+\eta\) reaction, aiming at determining the scattering length \(a\) and effective range \(r_0\) of the \(K^+\bar{K}^0\) interaction. These parameters are extracted by analyzing and fitting the mass distributions of the pairs in the final   \(\bar{K}^0\pi^+\eta\) state. To ensure the reliability of the results, we apply resampling techniques to evaluate statistical uncertainties and improve the precision of the scattering parameters. The obtained results are compared with previous theoretical predictions and experimental data, providing new insights into the \(K^+\bar{K}^0\) interaction at low energies.
\end{abstract}


\maketitle

\section{introduction}

The decay process \( D^+ \to K_s^0 \pi^+ \eta \) has been studied by the BESIII collaboration, first in Ref.~\cite{BESIII:2020pxp} and later with higher precision and an amplitude analysis in Ref.~\cite{BESIII:2023htx}. This process is useful for studying the \( a_0(980) \) resonance, which is an important scalar meson. The actual reaction is \( D^+ \to \bar{K}^0 \pi^+ \eta \), where the \( \bar{K}^0 \) is detected as a \( K_s^0 \) state.
{In Ref.~\cite{Ikeno:2024fjr}, the theoretical analysis of the  \( D^+ \to \bar{K}^0 \pi^+ \eta \) reaction was studied to understand the nature of the  \( a_0(980) \) in this reaction. The authors found that the  \( a_0(980) \) production is the dominant term in this reaction and also discussed other terms that interfere with this production mode.}

Initially, {as mentioned in Ref.~\cite{Ikeno:2024fjr},} this reaction is similar to the \( D^0 \to K^- \pi^+ \eta \) decay studied by Belle \cite{Belle:2020fbd}, as the latter can be obtained by replacing a \( \bar{d} \) quark with a \( \bar{u} \) quark.  However, there are notable differences. In the \( D^0 \to K^- \pi^+ \eta \) decay, the \( K^- \pi^+ \) system is dominated by the \( \bar{K}^{*0} \) resonance, creating a strong peak. In contrast, in the \( D^+ \to \bar{K}^0 \pi^+ \eta \) decay, the \( \bar{K}^0 \pi^+ \) system cannot originate from \( \bar{K}^{*0} \) or \( K^{*-} \), resulting in the absence of the \( \bar{K}^* \) contribution. This absence makes the process a good tool to study the \( a_0(980) \) resonance independently \cite{BESIII:2023htx}.

{The theoretical analysis of the \( D^0 \to K^- \pi^+ \eta \) reaction was studied in Ref. \cite{Toledo:2020zxj}}. It found that the \( K^- \pi^+ \) mass distribution has a sharp peak at the \( \bar{K}^{*0} \) threshold, while the \( \pi^+ \eta \)  mass distribution contains significant contributions of low and high invariant masses because of the \( \bar{K}^{*0} \) resonance. The \( K^- \eta \)  mass distribution also displays a double peak from the \( \bar{K}^{*0} \). On the other hand, in \( D^+ \to \bar{K}^0 \pi^+ \eta \), the \( \bar{K}^0 \pi^+ \)   mass distribution has no sharp structures, and the \( \pi^+ \eta \) distribution is dominated by the \( a_0(980) \) resonance,  which makes it easier to study this scalar meson. The \( \bar{K}^0 \eta \)  mass distribution also shows a significantly different behavior.

Studying the weak decays of heavy mesons is essential for understanding hadron interactions and dynamically generated states. Three-body decays are particularly useful   to explore hadronic final-state interactions and study the resonance nature of states \cite{Oller:1997ti,Kaiser:1998fi,Guo:2017jvc}. The \( D^+ \to \bar{K}^0 \pi^+ \eta \) reaction, which was  recently measured with high precision \cite{BESIII:2023htx}, provides an opportunity to investigate the \( a_0(980) \) resonance and extract the \( K^+\bar{K}^0 \) scattering parameters with minimal background interference from other factors.

The \( a_0(980) \) is known to strongly couple to the \( K\bar{K} \) channel and has been studied in various reactions \cite{Flatte:1976xu,Achasov:1987ts,Oller:1997ti,Kaiser:1998fi}.  However, it is difficult to determine the \( K^+\bar{K}^0 \) scattering length and effective range because of the complexity of final-state interactions and coupled-channel effects \cite{Buescher:2005uu,KLOE:2002kzf,Achasov:2002ir,E852:1996san,OBELIX:2002lhi}.  
 Sometimes, these parameters can be obtained from cusp structures in invariant mass distributions. This method has been successfully applied to the \( \pi^+ \pi^- \) system in \( K\to3\pi \) decays \cite{Colangelo:2006va,Batley:2009ubw} and more recently to the \( \Lambda_c^+ \to p K^- \pi^+ \) decay \cite{Belle:2022cbs}, which allowed a precise measurement of the \( \eta\Lambda \) scattering length \cite{Duan:2024okk}.

{In this work, following the study of Ref.~\cite{Ikeno:2024fjr},}  we focus on the \( D^+ \to \bar{K}^0 \pi^+ \eta \) reaction using a coupled-channel unitary approach \cite{Oller:1997ti,Kaiser:1998fi} to determine the \( K^+\bar{K}^0 \) scattering parameters.
This approach  has been successfully applied to studies of dynamically generated scalar mesons \cite{Oller:2000ma,Albaladejo:2016hae}. 
To get an accurate determination of the scattering parameters and their errors, we apply a statistical resampling method \cite{Efron:1986hys}, which helps to measure uncertainties in fits to the experimental data.

Previous experimental studies have provided estimates of the \( K^+\bar{K}^0 \) scattering length, but the results differ quite a bit \cite{Buescher:2005uu,KLOE:2002kzf,Achasov:2002ir,E852:1996san,OBELIX:2002lhi}. 
The high-statistics measurement by BESIII \cite{BESIII:2023htx} provide a unique opportunity to refine these values and improve our understanding of the strong interaction dynamics in the scalar sector. The results from this study will not only improve knowledge about the \( a_0(980) \) resonance but also provide  crucial input for future research on scalar mesons and their role in hadron spectroscopy \cite{Weinstein:1982gc,Baru:2003qq,Hyodo:2020czb}.

\section{Formalism}
\label{sec:formula}
{We study the $D^+ \to \bar K^0 \pi^+ \eta$ reaction and look at the mechanisms for external and internal emission based on the same formalism in Ref.~\cite{Ikeno:2024fjr}, except for the external emission process.}
\subsection{External emission}
\label{subsec:Ee}

\begin{figure}[H]
  \begin{center}
  \includegraphics[scale=0.48]{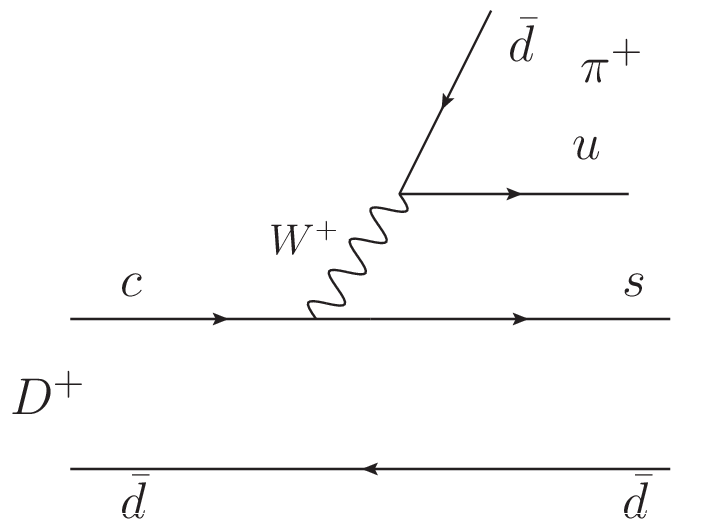}
  \end{center}
  \vspace{-0.5cm}
  \caption{Mechanism with external emission at the quark level. }
  \label{fig:Fig1}
\end{figure}

\begin{figure}[H]
\begin{center}
  \includegraphics[scale=0.48]{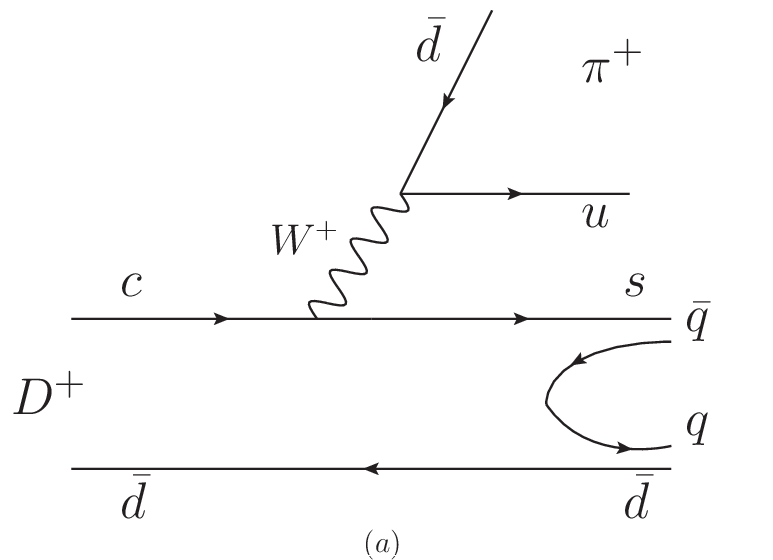}\qquad\qquad\qquad\qquad
  \includegraphics[scale=0.48]{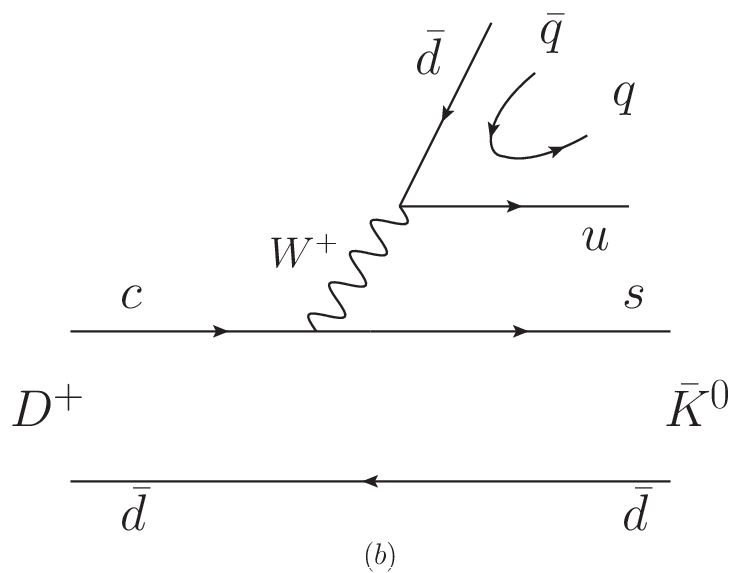}
\end{center}
\vspace{-0.5cm}
\caption{Hadronization of the  $s\bar d$  (a),  and hadronization of the  $u \bar d $ (b)  component  in external emission.}
\label{fig:Fig2}
\end{figure}

In Fig. \ref{fig:Fig1} we show the mechanism of external emission, Cabibbo favoured, at the quark level, and in  Fig. \ref{fig:Fig2} we show the hadronization of quark pairs into two mesons.
\noindent  In  Fig. \ref{fig:Fig2} (a), the $s\bar d$ pair is hadronized in the following way:
\begin{equation}
  s\bar d \to \sum_i s\;\bar q_i q_i\; \bar d,
\end{equation}
which is easily interpreted writing the $q_i \bar q_j$ matrix in terms of pseudoscalar mesons, $\mathcal{P}_{ij}$, which in the standard $\eta-\eta'$ mixing of Ref.~\cite{Bramon:1992kr} is given by

\begin{equation}\label{eq:Pmatrix}
  \mathcal{P} =
   \left(
   \begin{array}{cccc}
   \frac{1}{\sqrt{2}}\pi^0 + \frac{1}{\sqrt{3}}\eta  & \pi^+ & K^+  \\[2mm]
   \pi^- & -\frac{1}{\sqrt{2}}\pi^0 + \frac{1}{\sqrt{3}}\eta & K^0  \\[2mm]
   K^- & \bar{K}^0 & ~-\frac{1}{\sqrt{3}}\eta ~  \\
   \end{array}
   \right).
\end{equation}
Hence,
\begin{equation}\label{eq:sd}
  s\bar d \to \sum_i \;\mathcal{P}_{3i} \; \mathcal{P}_{i2}=\left( \mathcal{P}^2\right)_{32}=K^-\pi^+-\bar K^0 \,\dfrac{\pi^0}{\sqrt{2}},
\end{equation}
where the $\bar K^0 \eta$ component has canceled.

The combination of Eq.~\eqref{eq:sd}, together with the $\pi^+$ of Fig.~\ref{fig:Fig1}, does not lead to the desired final state $\bar K^0 \pi^+ \eta$.
However, through rescattering the $K^-\pi^+$ and $\bar K^0 \pi^0$ could lead to $\bar K^0 \eta$.
Yet, the threshold of $\bar K^0 \eta$ is $1040$~MeV, which is about $300$~MeV above the peak of the $K^*_0(700)$ (the kappa), where the $K^- \pi^+ \to \bar K^0 \eta$ amplitude has a reduced strength, and because of that we shall disregard this contribution, as it was also done in Ref.~\cite{Toledo:2020zxj}.

Next, with the same mechanism of Fig.~\ref{fig:Fig1}, we look at the hadronization of the $\bar d u$ component, as shown in Fig.~\ref{fig:Fig2}(b).
We have now
\begin{eqnarray}\label{eq:ud}
  u\bar d &\to  & \sum_i u \;\bar q_i q_i \;\bar d =\sum_i \;\mathcal{P}_{1i} \; \mathcal{P}_{i2}=\left( \mathcal{P}^2\right)_{12}.
\end{eqnarray}
Hence we have now the combination
\begin{equation}\label{eq:H1}
  H =
  (\frac{\pi^0}{\sqrt{2}} +\frac{\eta}{\sqrt{3}})\pi^+ + \pi^+(-\frac{\pi^0}{\sqrt{2}} +\frac{\eta}{\sqrt{3}}) +  K^+ \bar K^0.
\end{equation}
It might look like the two terms with $\eta\pi^+$ add and the two terms with $\pi^+\pi^0$ cancel. However, this is not the case as we show below. This is something peculiar of the external emission, when the $W^\mu$ couples directly to two mesons.
Indeed, as done in the Ref.~\cite{Ren:2015bsa},
one starts with the Langrangian of  $WPP$~\cite{Gasser:1983yg,Scherer:2002tk},
\begin{align}
    \mathcal{L}=\frac{f^2}{4}\left\langle D_\mu U\left(D^\mu U\right)^{\dagger}\right\rangle+\frac{f^2}{4}\left\langle\chi U^{\dagger}+U \chi^{\dagger}\right\rangle,
\end{align}
where
\begin{align}
&D_\mu U=\partial_\mu U-i r_\mu U+i U l_\mu,\nonumber\\
&U= \textrm{exp}(i\frac{\sqrt{2}P}{f}),
\end{align}
with $r_\mu,~l_\mu$
 the right-handed, left-handed weak interaction  currents
 \begin{align}
&r_\mu=v_\mu+a_\mu  = e Q A_\mu + \cdot\cdot\cdot, \nonumber\\
&l_\mu=v_\mu-a_\mu=e Q A_\mu+\frac{e}{\sqrt{2}\sin{\theta}_W}\left(W_\mu^{\dagger} T_{+}+W_\mu T_{-}\right),
\end{align}
where
\begin{equation}
T_{+}=\left(\begin{array}{ccc}
0 & V_{u d} & V_{u s} \\
0 & 0 & 0 \\
0 & 0 & 0
\end{array}\right), \quad T_{-}=\left(\begin{array}{ccc}
0 & 0 & 0 \\
V_{u d} & 0 & 0 \\
V_{u s} & 0 & 0
\end{array}\right),
\end{equation}
with $V_{u d},~V_{u s}$, matrix elements of the Cabibbo–Kobayashi–Maskawa matrix. 
Where, $v_\mu,~a_\mu,~A_\mu$, and $W_\mu$ are the vector current, axial current, electro magnetic field, and $W$ field,  $q$ the electron charge and $Q$ the matrix of the  charge of quarks $u,~d,~s$. We are only concerned on the weak part of $l_\mu$.
After expansion of the fields, one finds 
for the $WP_1P_2$ vertex of Fig. \ref{fig:Fig2} (b)
\begin{align}\label{eq:L}
\mathcal{L}=-i\frac{1}{2}\frac{e}{\sqrt{2}\sin{\theta}_W}W_\mu^+\langle [P,~\partial_\mu P] T_{+}\rangle.
\end{align}   
What matters for us here is 
the structure of the  term $ \langle [P,~\partial_\mu P] W^{\mu}\rangle  $. Indeed, if we have $\eta\pi^+$ and $\pi^+\eta$ terms, then we  get
\begin{align}
& \eta \partial_\mu \pi^+ - \pi^+\partial_\mu \eta,~~~\text{from the}~\eta\pi^+~\text{term};\nonumber\\
& \pi^+  \partial_\mu \eta- \eta \partial_\mu \pi^+,~~~\text{from the}~\pi^+\eta~\text{term},
\end{align}
and they cancel exactly.
By the contrary, the terms $\pi^0\pi^+$  and $\pi^+\pi^0$  of Eq.~(\ref{eq:H1}) add, but they do not contribute to the process. 
{The term $K^+\bar{K}^0$  of Eq.~(\ref{eq:H1}) could contribute through $K^+\bar{K}^0 \to \pi^+\eta$, but the structure of Eq.~(\ref{eq:L}), $K^+\partial_\mu \bar{K}^0 - \bar{K}^0 \partial_\mu K^+$, with the dominant $\mu=0$ component, also makes this term vanish in an average through the calculation of the difference of the $\bar{K}^0$ and $K^+$ energies.}  
In conclusion, we have found that there is no contribution to the process from external emission\footnote{We have carried out this detailed discussion, because in our previous paper of Ref.~\cite{Ikeno:2024fjr} this important fact was not realized, and there was a contribution to the mass distribution from external emission. Yet, the fit to the data demanded a contribution from external emission much smaller than from internal emission. The discussion done here, while supporting the numerical results of~\cite{Ikeno:2024fjr}, justifies this unusual conclusion in Ref.~\cite{Ikeno:2024fjr}. We take advantage to mention that the $V_{23}$ element of Eq.~(17) in \cite{Ikeno:2024fjr} should have an opposite sign. It is a typo and the right formula was used in the results.
}.

\subsection{Internal Emission}
\label{subsec:Ie}

\begin{figure}[H]
  \begin{center}
  \includegraphics[scale=0.48]{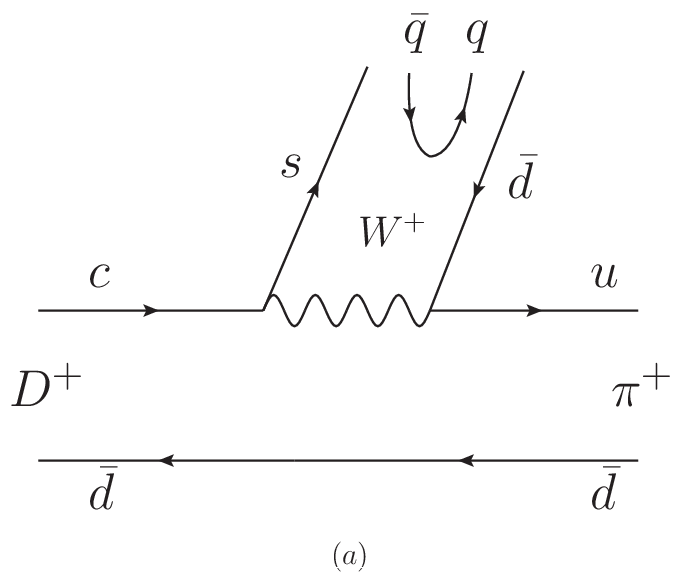}~~~~~~~~~~~~~~~~~~~~~~
  \includegraphics[scale=0.48]{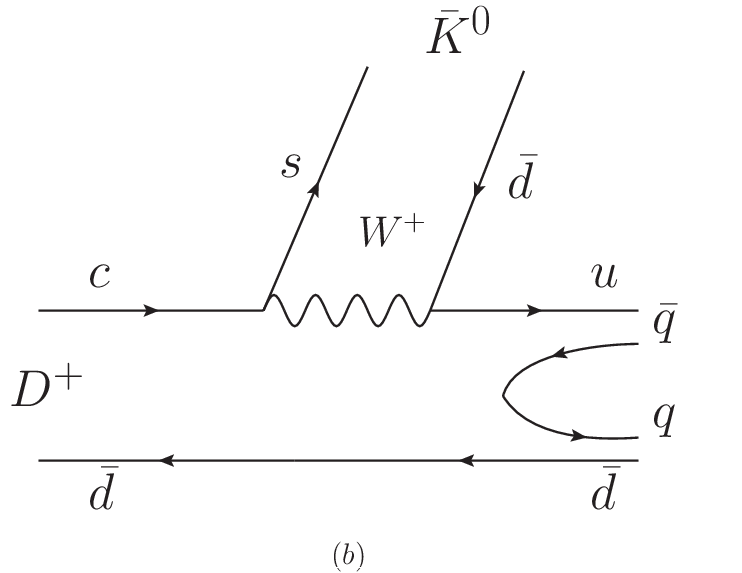}
  \end{center}
  \vspace{-0.5cm}
  \caption{Diagrams representing internal emission: (a) hadronization of the $s\bar d$ pair; (b) hadronization of the $u\bar d$ pair.}
  \label{fig:Fig4}
  \end{figure}

{For internal emission, we use the same formalism of Ref.~\cite{Ikeno:2024fjr}, and we show the same formula here for completeness.} We utilize the diagrams in Fig.~\ref{fig:Fig4}, which depict the hadronization of quark pairs.

Following the process outlined in subsection \ref{subsec:Ee}, we now describe the hadronization:

\begin{eqnarray}\label{eq:sd2}
  s\bar d &\to& \sum_i s \;\bar q_i q_i \;\bar d =\sum_i \;\mathcal{P}_{3i} \; \mathcal{P}_{i2}=\left( \mathcal{P}^2\right)_{32} = K^- \pi^+ -\dfrac{1}{\sqrt{2}}\, \bar K^0 \pi^0 ,
\end{eqnarray}
where the $\bar K^0 \eta$ terms cancel, and

\begin{eqnarray}\label{eq:ud2}
  u\bar d &\to& \sum_i u \;\bar q_i q_i \;\bar d =\sum_i \;\mathcal{P}_{1i} \; \mathcal{P}_{i2}=\left( \mathcal{P}^2\right)_{12} = \dfrac{2}{\sqrt{3}}\, \eta \pi^+ + K^+ \bar K^0,
\end{eqnarray}
where the $\pi^+ \pi^0$ terms cancel.

\begin{figure}[H]
  \begin{center}
  \includegraphics[scale=0.4]{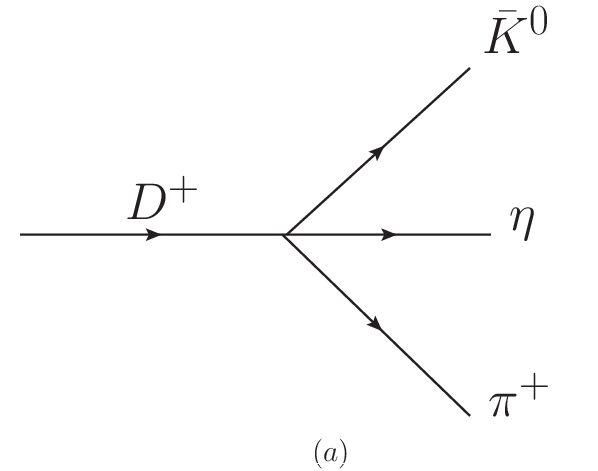}~~~~~~~~~~~~~~~~~~~~~~~~~~~~~~~~~~~~
  \includegraphics[scale=0.45]{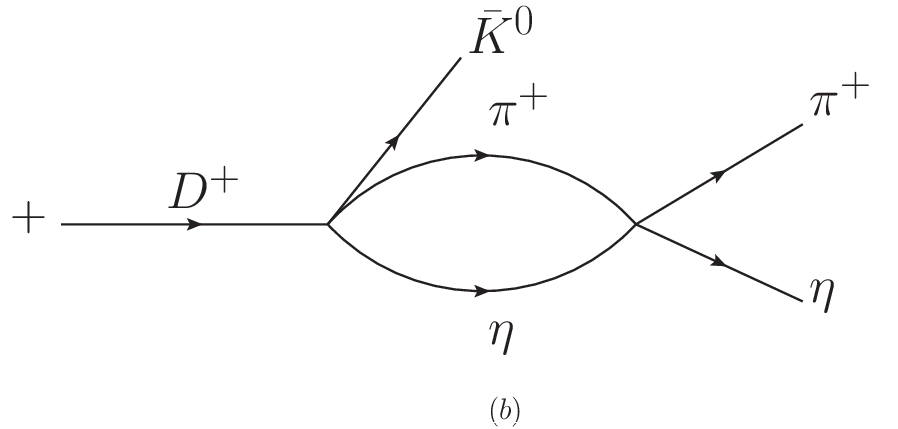}\\
  \includegraphics[scale=0.45]{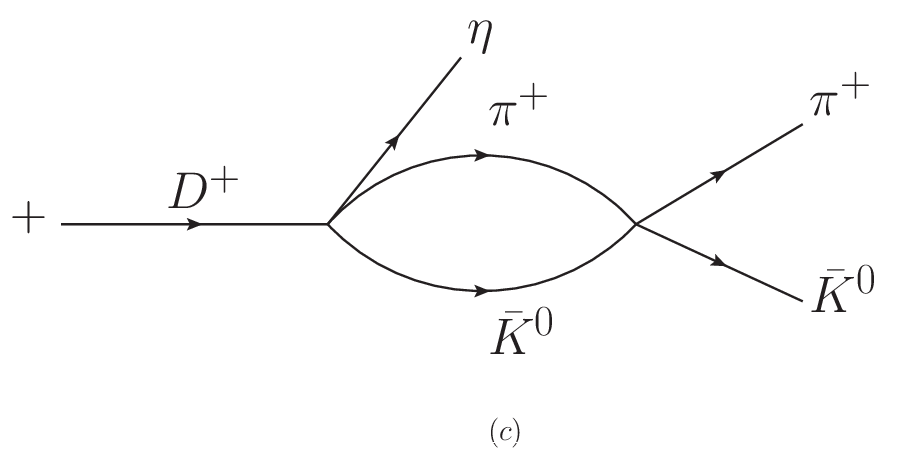}  \includegraphics[scale=0.45]{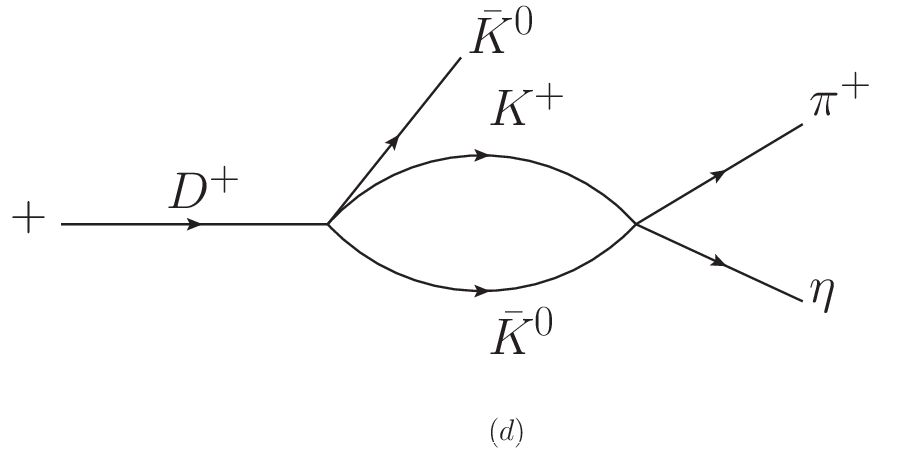}  \end{center}
  \vspace{-0.5cm}
  \caption{Diagrams showing the production of $\eta \pi^+ \bar K^0$ from internal emission: (a) tree level; (b) $\pi^+ \eta$ rescattering; (c) $\pi^+ \bar K^0$ rescattering; (d) $K^+ \bar K^0$ rescattering.}
  \label{fig:Fig5}
  \end{figure}

Summing the two terms, including the $\pi^+$ from Fig.~\ref{fig:Fig4} (a) and $\bar K^0$ from Fig.~\ref{fig:Fig4} (b), we obtain:

\begin{equation}\label{eq:H'}
  H'=K^- \pi^+ \pi^+ -\dfrac{1}{\sqrt{2}}\; \pi^0 \pi^+ \bar K^0 + \dfrac{2}{\sqrt{3}}\; \eta \pi^+ \bar K^0 + K^+ \bar K^0 \bar K^0.
\end{equation}
Once again we have a tree-level production of $\eta \pi^+ \bar K^0$, with the other terms potentially leading to this final state via rescattering, as shown in Fig.~\ref{fig:Fig5}.

In Fig.~\ref{fig:Fig5}, we have ignored possible rescattering processes such as $K^- \pi^+ \to \bar K^0 \eta$, $\bar K^0 \pi^0 \to \bar K^0 \eta$, and $\bar K^0 \eta \to \bar K^0 \eta$, for the same reasons discussed in the previous subsection. Numerical checks confirm that these terms are negligible.

We analytically describe the diagrams of Fig.~\ref{fig:Fig5} corresponding to internal emission as:
\begin{align}\label{eq:tie}
 t^{(ie)} = & \mathcal{C}~ \Bigg\{ h_{\eta\pi^+\bar{K}^0} \Bigg[ 1 + G_{\pi^+\eta}\Bigg(M_\text{inv}(\pi^+\eta)\Bigg)~t_{\pi^+\eta, \pi^+\eta}\Bigg(M_\text{inv}(\pi^+\eta)\Bigg)
 + G_{\pi^+\bar{K}^0}\Bigg(M_\text{inv}(\pi^+\bar{K}^0)\Bigg)~t_{\pi^+\bar{K}^0, \pi^+\bar{K}^0}\Bigg(M_\text{inv}(\pi^+\bar{K}^0)\Bigg)\Bigg]~~\nonumber\\ 
 &+ 2  h_{K^+\bar{K}^0\bar{K}^0}  G_{K^+\bar{K}^0}\Bigg(M_\text{inv}(\pi^+\eta)\Bigg)~t_{K^+\bar{K}^0, \pi^+\eta}\Bigg(M_\text{inv}(\pi^+\eta)\Bigg) \Bigg\},
\end{align}
where the weights $h_i$ are given by:

\begin{equation}
   h_{\eta \pi^+ \bar K^0}= \dfrac{2}{\sqrt{3}};~~~
   h_{K^+ \bar K^0 \bar K^0}=1,
\end{equation}
and $\mathcal{C}$ is a global normalization factor to be fitted to the normalization of the data.
The factor of 2 multiplying $ h_{K^+ \bar K^0 \bar K^0}$ in Eq.~\eqref{eq:tie} accounts for the two identical $\bar K^0$ particles.

\subsection{$K^*_0(1430)$ Contribution}
\label{subsec:K0}
\begin{figure}[H]
  \begin{center}
  \includegraphics[scale=0.66]{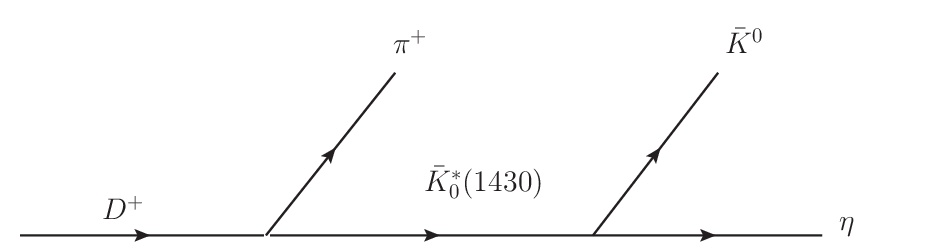}
  \end{center}
  \vspace{-0.5cm}
  \caption{Diagram representing the process $D^+ \to \pi^+ \bar K^*_0(1430) \to \pi^+ \bar K^0 \eta$.}
  \label{fig:Fig6}
  \end{figure}

{We follow the same formula of Ref.~\cite{Ikeno:2024fjr} to consider the $K^*_0(1430)$ resonance, where the  important role of   $K^*_0(1430)$  was discussed.}
In Ref.~\cite{BESIII:2023htx}, it was observed that the scalar $K^*_0(1430)$ [$I\, (J^P)=\frac{1}{2}\, (0^+)$] state contributes to the $D^+  \to  \bar K^0 \pi^+ \eta$ process and appears in the $K\eta$ mass distribution. The process is depicted in Fig.~\ref{fig:Fig6} (we label the particles as $\bar K^0$ (1), $\pi^+$ (2), $\eta$(3)).

We account for this contribution phenomenologically by introducing the amplitude:

\begin{equation}\label{eq:K0star}
  t^*= \mathcal{D} \, e^{i \phi}\; \dfrac{M_D^2}{s_{13}-M^2_{K^*_0}+i M_{K^*_0}\, \Gamma_{K^*_0}},
\end{equation}
where $\mathcal{D}$ and $\phi$ are free parameters, and $s_{13}=(p_{\bar K^0}+p_\eta)^2$ represents the invariant mass of the $\bar K^0$ and $\eta$ particles. The factor $M_D^2$ is included to make $\mathcal{D}$ dimensionless.
We take the mass and width from the PDG~\cite{ParticleDataGroup:2024cfk} as $ M_{K^*_0} = 1425$ MeV, $\Gamma_{K^*_0} = 270$ MeV.

The total amplitude now becomes:
\begin{equation}\label{eq:ttotal2}
  t=  t^{(ie)} + t^*.
\end{equation}
To calculate the mass distributions, we use the standard PDG formula \cite{ParticleDataGroup:2024cfk}:
\begin{equation}\label{eq:Gamm}
  \dfrac{d^2 \Gamma}{ds_{12}\; ds_{23}} = \dfrac{1}{(2\pi)^3}\; \dfrac{1}{32\, M^3_D}\; |t|^2.
\end{equation}
By using  the limits of the PDG formula \cite{ParticleDataGroup:2024cfk}, we can obtain $d \Gamma / ds_{12}$ by integrating over $s_{23}$. Using cyclical permutations of the formulas, we can derive $d\Gamma / ds_{13}$ and $d \Gamma / ds_{23}$ and compare with experimental data \cite{BESIII:2023htx}.

\subsection{Scattering Amplitudes}
\label{subsec:tij}

We need the following amplitudes:

\begin{equation*}
  t_{\eta \pi^+, \, \eta \pi^+}, ~~ t_{\bar K^0 \pi^+, \, \bar K^0 \pi^+}, ~~ t_{\bar K^0 K^+, \, \pi^+ \eta}.
\end{equation*}
Using the results from Ref.~\cite{Lin:2021isc} (Eq.~(A.4) of Ref.~\cite{Lin:2021isc}), which explicitly considers $\eta-\eta'$ mixing, we find the matrix elements of the potential for the channels $K^+ K^- (1), K^0 \bar K^0 (2), \pi^0 \eta (3)$:

\begin{alignat}{2}\label{eq:Vij}
& V_{11}=-\dfrac{s}{2f^2},        &\quad \quad &V_{12}= -\dfrac{s}{4f^2}, \nonumber\\[1mm] 
& V_{13}= -\dfrac{3s-2m^2_K-m^2_\eta}{3\sqrt{6}f^2}, & \quad \quad & V_{22}=-\dfrac{s}{2f^2},        \nonumber\\[1mm]
& V_{23}= \dfrac{3s-2m^2_K-m^2_\eta}{3\sqrt{6}f^2}, &\quad \quad &V_{33}= -\dfrac{2m^2_\pi}{3f^2}.
\end{alignat}
Then, the amplitudes $t$ are obtained from the Bethe-Salpeter equation in coupled channels:
\begin{equation}\label{eq:BS}
  T= [1-VG]^{-1}\, V.
\end{equation}
Considering that $|\pi^+\rangle =-|11\rangle$ of isospin, we find

\begin{equation}\label{eq:t1}
  t_{\eta \pi^+, \, \eta \pi^+} = t_{\eta \pi^0, \, \eta \pi^0}.
\end{equation}
Also, taking into account that $K^+K^-$ in terms of $|I,\, I_3\rangle$ is:

\begin{equation}
  K^+K^- = - \left( \dfrac{1}{\sqrt{2}} \, |00\rangle+\dfrac{1}{\sqrt{2}} \, |10\rangle \right),
\end{equation}
then we obtain the $\bar K^0 K^+ \to \, \pi^+ \eta$ amplitudes
\begin{equation}\label{eq:t2}
  t_{\bar K^0 K^+, \, \pi^+ \eta} = \sqrt{2}\; t_{K^+ K^-,\, \pi^0 \eta}.
\end{equation}
We also require the $K \pi \rightarrow K \pi$ amplitudes, which we extract from the Appendix of Ref.~\cite{Toledo:2020zxj}, considering the channels $\pi^{-} K^{+}(1)$, $\pi^0 K^0(2)$, and $\eta K^0(3)$. By applying isospin coefficients and the $C$-parity relation $C \bar{K}^0 \pi^{+} = K^0 \pi^{-}$, we obtain
\begin{align}\label{eq:t3}
t_{\bar{K}^0 \pi^{+}, \bar{K}^0 \pi^{+}}=\frac{2}{3} T_{22}+\frac{1}{3} T_{11}+\frac{2 \sqrt{2}}{3} T_{12}.     
\end{align}
And the loop function $G_l$ in Eq.~(\ref{eq:tie}) is given by:
\begin{align}\label{cut}
   G_l(\sqrt{s}) = \int_{|q|<q_{\text{max}}} \frac{d^3 q}{(2 \pi)^3} \frac{ \left( w_1(q) + w_2(q) \right)}{2 w_1(q) w_2(q)} \frac{1}{s - (w_1(q) + w_2(q))^2 + i \epsilon},
\end{align}
where \( w_i(q) = \sqrt{m_i^2 + \vec{q~}^2},~i=1,2 \), with $1,~2$ referring to the two particles of channel $l$. The cutoff is \( q_{\text{max}} = 630~\text{MeV} \)~\cite{Lin:2021isc}.
\subsection{Determination of $a$  and $r_0$ for $K^{+}\bar{K}^{0}$}
The scattering length \( a \) and effective range \( r_0 \) are given in Ref.~\cite{Ikeno:2023ojl},
\begin{align}  
-\frac{1}{a} &= -\left.8 \pi \sqrt{s} T^{-1}\right|_{s=s_{\mathrm{th}}}, \nonumber\\  
r_0 &= \left.\frac{\sqrt{s}}{\mu} \frac{\partial}{\partial s} 2\left(-8 \pi \sqrt{s} T^{-1}+i k\right)\right|_{s=s_{\mathrm{th}}}, \nonumber\\  
k =& \frac{\lambda^{1 / 2}\left(s, m_{K^+}^2, m_{\bar{K}^0}^2\right)}{2 \sqrt{s}},
\end{align}  
where $T$ stands for the $T_{K^{+}\bar{K}^{0},K^{+}\bar{K}^{0}}$  scattering amplitude, \( s_{\text{th}} \) is the squared energy of the 
$K^{+}\bar{K}^{0}$ system at threshold, and \( \mu \) is the reduced mass of \( {K^+} \) and \( {\bar{K}^0} \).  
To evaluate \( a \) and \( r_0 \) for \( K^{+}\bar{K}^{0} \), one must have very precise values for the thresholds.  
 Since we only need \( I=1 \) in this case,  
we can directly use the \( K^+ \bar K^0 \) and \( \pi^+\eta \) channels.  

The \( V \)-matrix for the coupled channels \( K^+ \bar{K}^0 \) and \( \pi^+\eta \) is given by  
\begin{align}  
V_{11} &= -\frac{s}{4 f^2},~~  
V_{12} = -\frac{3 s - m_{K^+}^2 - m_{\bar{K}^0}^2 - m_\eta^2}{3 \sqrt{3} f^2}, \nonumber\\  
V_{22} &= -\frac{2 m_\pi^2}{3 f^2},  
\end{align}  
with \( f=93 \)~MeV.  
The \( T \) matrices are then obtained using the Bethe-Salpeter equation in Eq.~\eqref{eq:BS}.

\section{Results}
\subsection{ Mass distributions of $\bar{K}^0 \pi^+$, $\bar{K}^0 \eta$, and $\pi^+ \eta$}
Our theoretical model involves $3$ parameters: $\mathcal{C}$,  $\mathcal{D}$, and $\phi$. To determine their values, we perform a best fit to the mass distributions of $\bar{K}^0 \pi^+$, $\bar{K}^0 \eta$, and $\pi^+ \eta$ using the experimental data from Ref.~\cite{BESIII:2023htx}. The resulting parameter values are listed in Table~\ref{tab:param}. 
\begin{table}[H]
\centering
\caption{ The value of the parameters $C$, $D$, and $\phi$ [radians] from the best fit. }
\label{tab:param}
\setlength{\tabcolsep}{68pt}
\begin{tabular}{ccc}
\hline \hline
$C$ & $D$ & $\phi$ [radians] \\
\hline 
$ 592.48 $ & $51.328 $	 & $-1.747 $\\
\hline \hline
\end{tabular}
\end{table}
The parameter $\mathcal{C}$ serves as an overall normalization factor.  The parameter $\mathcal{D}$ determines the strength of the $K_0^*(1430)$ contribution. Moreover, the phase $\exp(i \phi)$ accounts for the interference between the $K_0^*(1430)$ and other components, playing a important role in shaping the mass distributions.

\begin{figure}[H]
  \centering
\includegraphics[width=0.33\textwidth]{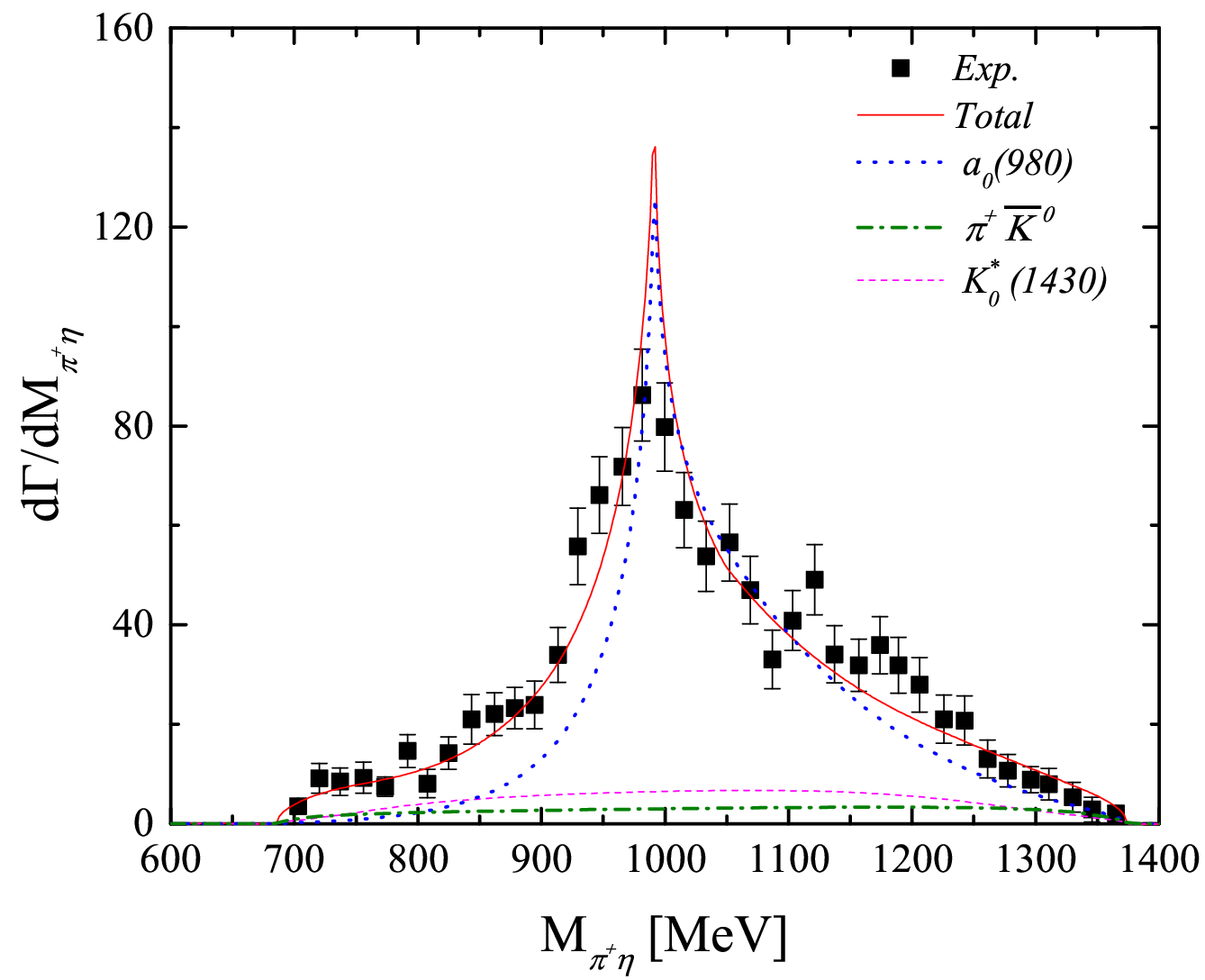}
\includegraphics[width=0.33\textwidth]{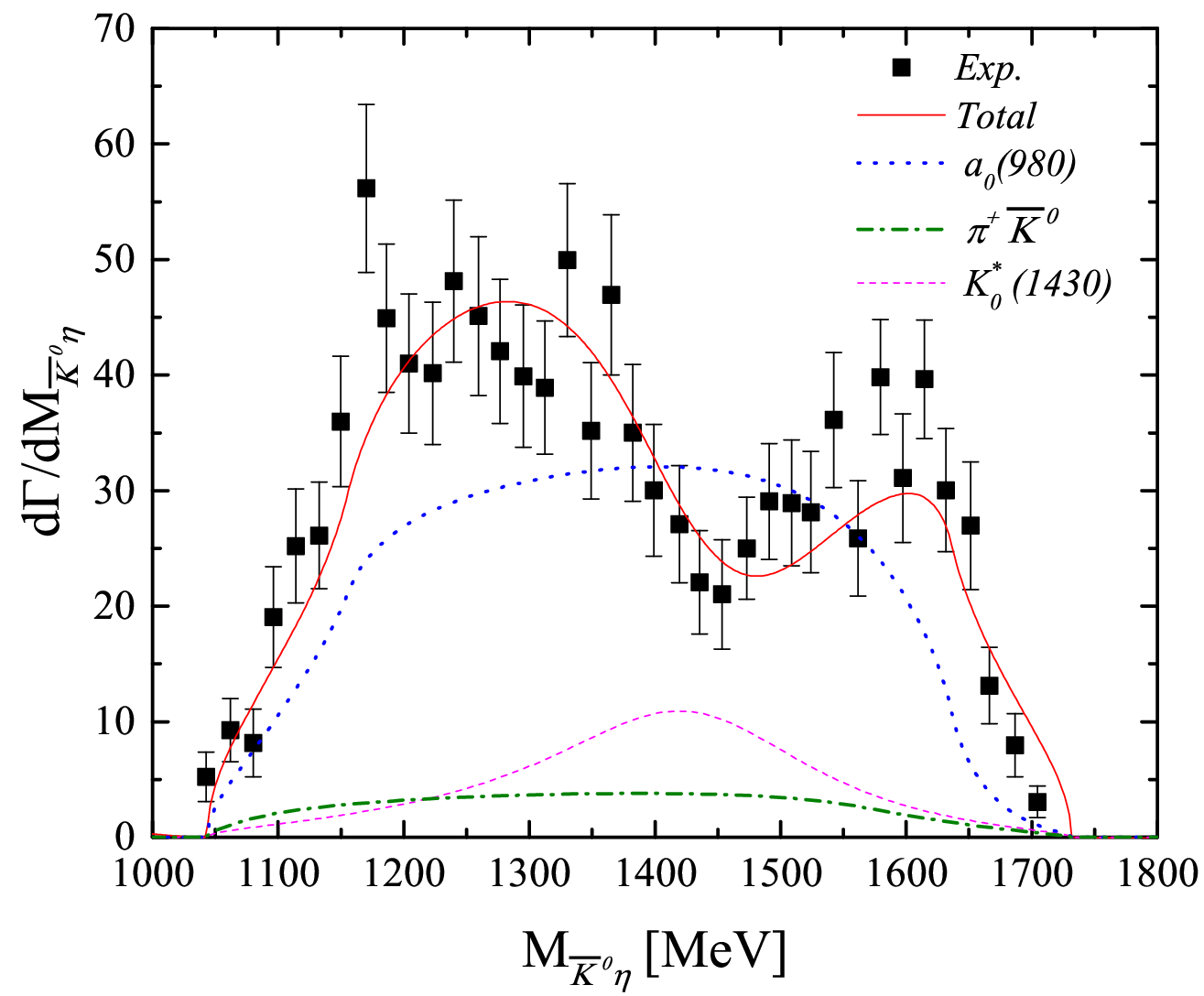}
\includegraphics[width=0.33\textwidth]{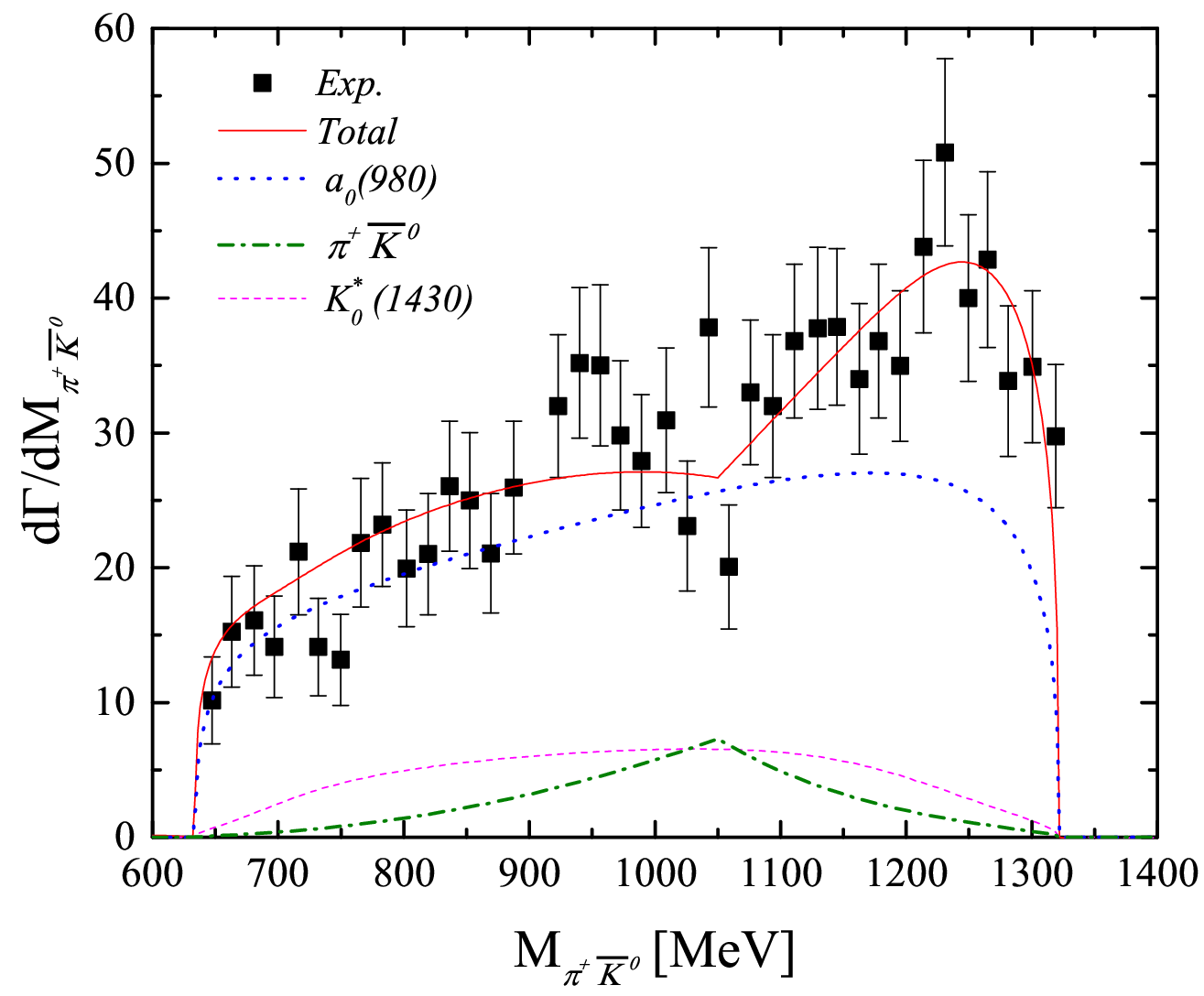} 
  \caption{The mass distributions of $\pi^+ \eta$ (left), $\bar{K}^0 \eta$ (middle), and $\bar{K}^0 \pi^+$ (right) with fixed $M_{\text{cut}} = 1050$ MeV. The parameters $\mathcal{C} = 592.48$, $\mathcal{D} = 51.328$,  and $\phi = -1.747$ radians are used .}\label{fig:dGdM_1}
\end{figure}

Figure~\ref{fig:dGdM_1} presents the mass distribution results obtained with the fitted parameters from Table~\ref{tab:param}. A comparison with the experimental data shows that our theoretical predictions accurately describe well the observed distributions.
{We found that the features of the results obtained here in the revised formalism are the same as those in Ref.~\cite{Ikeno:2024fjr}, which were discussed in detail in Ref.~\cite{Ikeno:2024fjr}. We  recall the main  features here.}
In Fig.~\ref{fig:dGdM_1} (left), a pronounced peak appears in the $\pi^+\eta$ spectrum around 1.0 GeV, corresponding to the $a_0(980)$ resonance. In our model, this resonance is encoded in the $T$-matrix elements $t_{\eta\pi^+,\,\eta\pi^+}$ and $t_{\bar{K}^0 K^+,\, \eta\pi^+}$, as defined in Eqs.~\eqref{eq:t1} and \eqref{eq:t2}. On the other hand, the contribution from the $\bar{K}^0\pi^+$ scattering term in Eq.~\eqref{eq:t3} and the $K_0^*(1430)$ component is significantly smaller than that of the $a_0(980)$ resonance. 
It is also important to emphasize that the interference with other components, particularly the tree-level contributions in Eqs.~\eqref{eq:tie}, leads to a broader lineshape of the $a_0(980)$ resonance compared to other processes. 

Figure~\ref{fig:dGdM_1} (middle) displays the $\bar{K}^0 \eta$ mass distribution, where a characteristic double-hump structure appears. The $K^*_0(1430)$ resonance contributes a peak at $M_{\bar K^0 \eta} = M_{K^*_0} = 1425$ MeV with a width of $\Gamma_{K^*_0} = 270$ MeV, as given in Eq.~\eqref{eq:K0star}. 
{As discussed in the detail in Ref.~\cite{Ikeno:2024fjr},} the interference between the $K^*_0(1430)$ and $a_0(980)$ components is responsible for the double-hump shape, requiring the phase $\phi$ in Table~\ref{tab:param} to be negative.

Figure~\ref{fig:dGdM_1} (right) shows the $\bar{K}^0\pi^+$ mass distribution, which lacks a distinct peak but exhibits a broad bump around 1.25 GeV. Additionally, a discontinuity is visible at approximately 1.05 GeV. {As discussed in the detail in Ref.~\cite{Ikeno:2024fjr},} this feature results from applying a cut-off mass $M_{\textrm{cut}}$ to the $G\cdot t$ terms in Eq.~\eqref{eq:tie}, 
\begin{equation}
    Gt(M_\mathrm{inv})=Gt(M_\mathrm{cut})\,e^{-\alpha\,(M_\mathrm{inv}-M_\mathrm{cut})}
\end{equation}
following the approach outlined in Refs.~\cite{Debastiani:2016ayp,Toledo:2020zxj} to avoid going outside the range of validity of the chiral unitary approach. 
 In this study, we use $M_{\textrm{cut}} = 1050$ MeV and $\alpha = 0.0037$ MeV$^{-1}$.

\subsection{ Evaluation of $a$, $r_{0}$ }
{In this part, we address the issue of evaluating the scattering length and effective range, not addressed in Ref.~\cite{Ikeno:2024fjr}, which is the main purpose of the work.}
First we show the results for $a$ and $r_{0}$ using the chiral unitary approach, with the potentials employed, following Refs.~\cite{Liang:2014tia,Xie:2014tma} with $f=93$ MeV as described before. The results are shown in Table~\ref{tab:1}.
These values are consistent with those of Ref.~\cite{Li:2024uwu}.
\begin{table}[H]
\centering
\caption{ The value of  $a$ and $r_0$[units in $fm$]. }
\label{tab:1}
\setlength{\tabcolsep}{88pt}
\begin{tabular}{ccc}
\hline \hline
 $a_{K^+\bar{K}}$ & $r_{0,K^+\bar{K}}$\\
 \hline 
 $-0.124 -i~ 0.741$ & $ -0.782 -i~ 0.182	 $ \\
\hline \hline
\end{tabular}
\end{table}

\subsection{Resampling for $a$ and $r_{0}$ with uncertainties}

\begin{figure}[H]
  \centering
\includegraphics[width=0.33\textwidth]{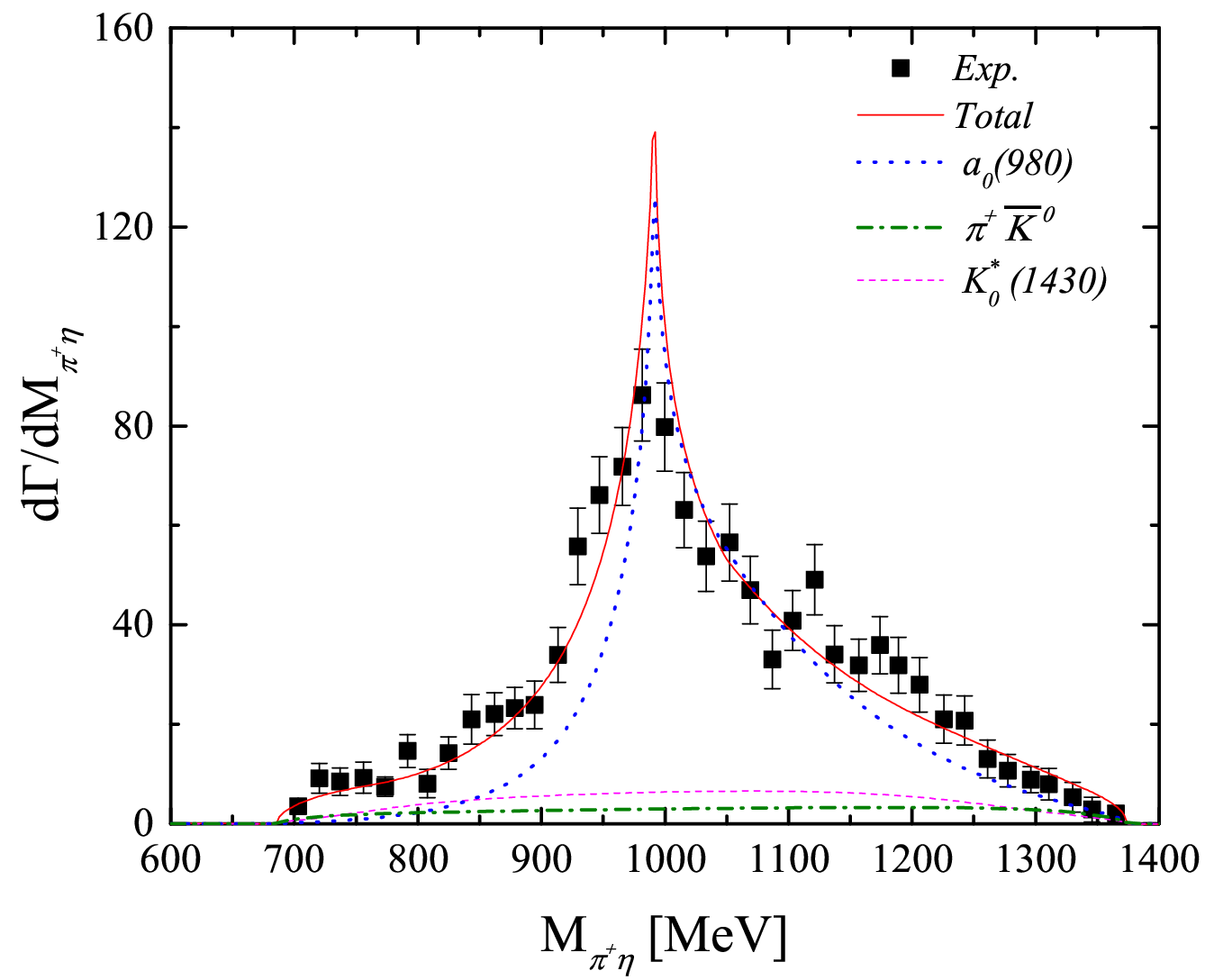}
\includegraphics[width=0.33\textwidth]{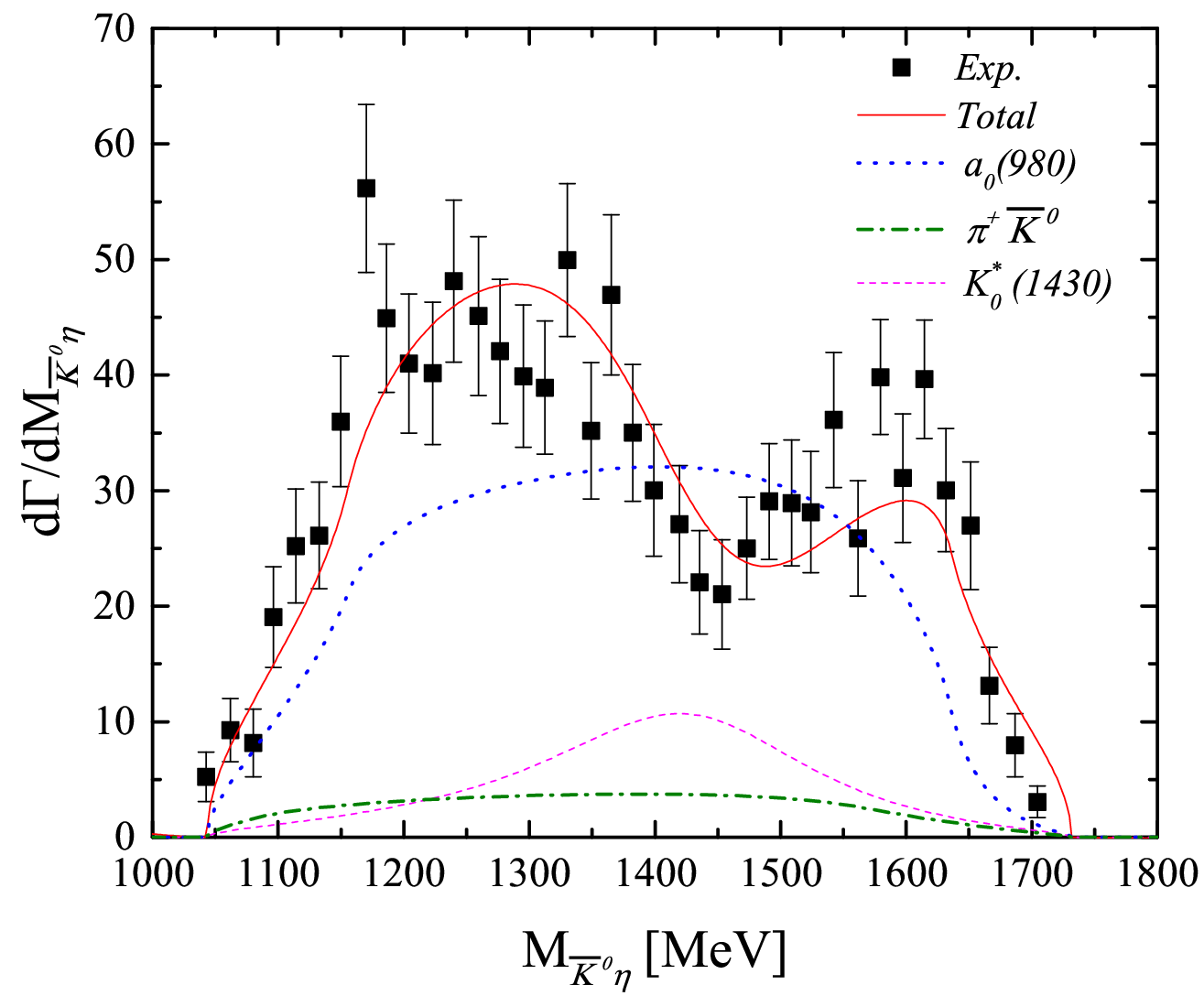}
\includegraphics[width=0.33\textwidth]{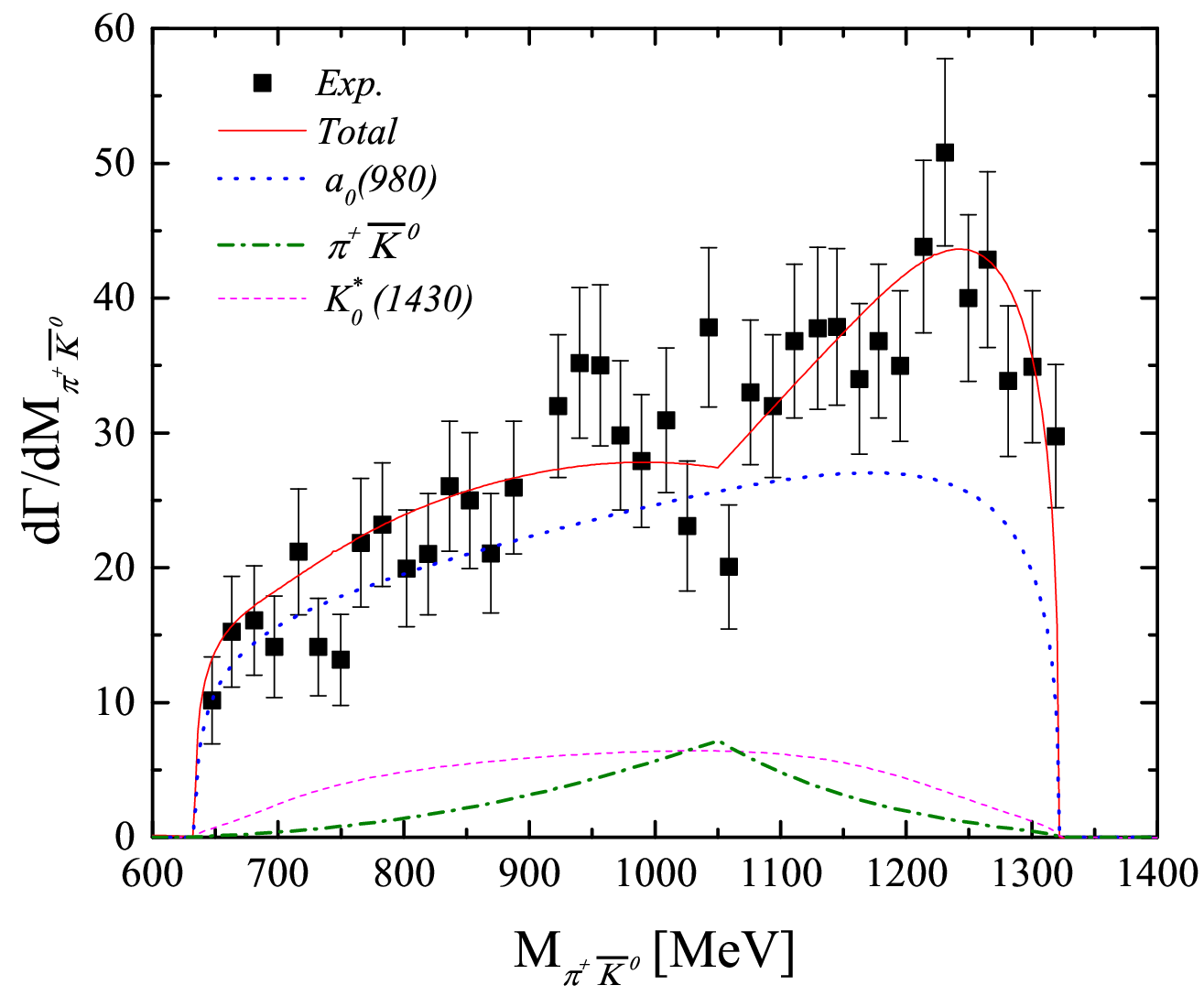} 
  \caption{After resampling: The mass distributions of $\pi^+ \eta$ (left), $\bar{K}^0 \eta$ (middle), and $\bar{K}^0 \pi^+$ (right) with fixed $M_{\text{cut}} = 1050$ MeV. The parameters $\mathcal{C} = 596.03$, $\mathcal{D} = 49.86$,  and $\phi =  -1.85 $ radians are used.}\label{fig:dGdM_2}
\end{figure}

To estimate uncertainties in $a$ and $r_{0}$, we employ a resampling approach~\cite{Press:1992zz,Efron:1986hys,Albaladejo:2016hae}, fitting the parameters of the theory to data with Gaussian randomly distributed centroids of the points, with the same error. The procedure follows these steps: all terms in the interaction kernel $V_{ij}$ from Refs.~\cite{Liang:2014tia,Xie:2014tma} are proportional to $\frac{1}{f^2}$, where $f$ is the pion decay constant. In Ref.~\cite{Liang:2016hmr}, $f=93$ MeV was used. Here, to allow flexibility in the potential and perform a model-independent analysis, we introduce the following modifications:
\begin{align}
\nonumber
f\to& f_{\pi\eta},& \text{in the $\pi\eta$ channel};\\
\nonumber
f\to& f_{K\bar{K}},& \text{in the $K\bar{K}$ channel}.
\end{align}
Then we obtain  the different $V-$matrix of the coupled channels ${K^+}{\bar{K}^0}(1)$ and $\pi^+\eta(2)$
\begin{align}
V_{11}=-\frac{s}{4 f_{K\bar{K}}^2},~~ 
&V_{12}=-\frac{3 s- m_{K^+}^2-m_{\bar{K}^0}^2-m_\eta^2}{3 \sqrt{3} _{K\bar{K}}f_{\pi\eta}} \nonumber\\
& V_{22}=-\frac{2 m_\pi^2}{3 f_{\pi\eta}^2}. 
\end{align}
Now, we have 6 parameters $C,~D,~\phi,~f_{K\bar{K}},~f_{\pi\eta},~q_\mathrm{max}$
where the values of $f_i$ are sampled within the range $[40,~180]$ MeV.
Additionally, we adjust the cutoff parameter, sampling $q_{\text{max}}$ within the range $[400,~1500]$ MeV. The resulting values of $C,~D,~\phi,~f_i$, $a_{K^{+}\bar{K}^{0}}$, $r_{0,K^{+}\bar{K}^{0}}$, and $q_{\text{max}}$ are presented in Table~\ref{tab:2}.
Using these parameters, we plot the different mass distributions in Fig.~\ref{fig:dGdM_2}. A comparison with the experimental data shows that our theoretical results are in agreement with the data.

\begin{table}[H]
\centering
\caption{After resampling 100 times $\chi^2=2.25/d.o.f$ as a average for the different fits. }
\label{tab:2}
\setlength{\tabcolsep}{34pt}
\begin{tabular}{ccc}
\hline \hline
$C$ & $D$ & $\phi$  [radians]  \\
\hline 
$596.03 \pm 29.92 $ & $ 49.86 \pm  7.28 $	 & $ -1.85  \pm 0.64  $\\
\hline 
$f_{K^+\bar{K}}$ [MeV] &  $f_{\pi\eta}$ [MeV] &$q_{\text {max }}$ [MeV]\\
\hline 
 $ 93.69  \pm 5.88  $  &	$ 92.96\pm 5.81 $ & $ 622.27 \pm 39.31  $\\
\hline \hline
 $a_{K^+\bar{K}^0}$ [fm]& $r_{0,K^+\bar{K}^0}$ [fm]\\
 \hline 
 ($ -0.10  \pm0.15)  -i~  (0.70 \pm 0.14 ) $ &  $(-0.81 \pm 0.15)  -i~  (0.20 \pm 	0.09) $ \\
\hline \hline
\end{tabular}
\end{table}
These results indicate that $\text{Re} [a] \approx -0.10$ fm, while $\text{Im}  [a] \approx -0.70$ fm, with uncertainties that are consistent with Table~\ref{tab:1}. Similarly, $\text{Re} [r_{0}] \approx -0.81$ fm and $\text{Im} [r_{0}] \approx -0.20$ fm.

\begin{table}[H]
\centering
\caption{Results for the $K^+\bar{K}^0$  scattering length from various studies. }
\label{tab:new}
\setlength{\tabcolsep}{8pt}
\begin{tabular}{cccccc}
\hline \hline
{ References } &  { Ref. \cite{KLOE:2002kzf} } &  { Ref. \cite{Achasov:2002ir} } &  { Ref. \cite{Achasov:2000ku} } &  { Ref. \cite{Bugg:1994mg} } &   { Ref. \cite{Buescher:2005uu} } \\
\hline 
$a(\mathrm{fm})$ & $+0.071-i~0.66 $ & $-0.006-i~0.76 $ & $-0.13-i~2.23 $ & $-0.075-i~0.70 $ &$ (-0.02 \pm 0.02)-i~(0.63 \pm 0.24) $ \\
\hline 
 { References } &  { Ref. \cite{Achasov:2002ir} } &  { Ref. \cite{Abele:1998qd} } &  { Ref. \cite{E852:1996san} } &  { Ref. \cite{OBELIX:2002lhi} } &  Ref. \cite{Li:2024uwu}  \\
\hline 
$a(\mathrm{fm})$ & $-0.16-i~1.05 $ & $-0.13-i~0.61 $ & $-0.16-i~0.59 $ & $-0.54-i~1.89 $ & $-0.371-i~0.549$ \\
\hline \hline
\end{tabular}
\end{table}
It is interesting to compare those results with other values obtained from different approaches and reactions for $a_{K^+\bar{K}^0}$, which we present in Table~\ref{tab:new}. As we can see, there is quite some dispersion in the values given by the different approaches. Our determination of $a_{K^+\bar{K}^0}$ provides also the uncertainty, and it is based on the analysis of a reaction with quite good statistics and a clean signal for the $a_0(980)$. In particular we would like to compare our results with those of~\cite{Li:2024uwu}, which were determined in a similar way as done here, with a resampling analysis of the data of~\cite{BESIII:2016tqo} for the $\chi_{c1} \to \eta\pi^+\pi^-$ reaction.
The results obtained in~\cite{Li:2024uwu} are 
\begin{align}\label{eq:new}
 & a_{K^+\bar{K}^0}=  -(0.371 \pm 0.045)-i(0.549 \pm 0.023), \nonumber\\
 & r_{0,K^+\bar{K}^0} = -(0.982 \pm 0.107)-i(0.265 \pm 0.035).
\end{align}
We first see that the errors in Eq.~(\ref{eq:new}) are smaller than those that we get here in Table~\ref{tab:2}. This is understandable since the experimental errors in~\cite{BESIII:2016tqo} are smaller than in the reaction studied here~\cite{BESIII:2023htx}. If we look at $\text{Re}(a_{K^+\bar{K}^0})$, playing with uncertainties, the results of~\cite{Li:2024uwu} would provide a minimum value of $-0.32$~fm (minimum of the absolute value), while the maximum value here is $-0.26$~fm. They are not far away, and   the difference is small when compared with the dispersion of values 
in Table~\ref{tab:new}. So, we can give a more precise value of $\text{Re} (a_{K^+\bar{K}^0})$ than  can be  anticipated   from the values in Table~\ref{tab:new}. On the other hand $\text{Im}(a_{K^+\bar{K}^0}) =-( 0.549 \pm 0.023)$~fm in~\cite{Li:2024uwu} overlaps within errors with our value in Table~\ref{tab:2}.

We take advantage to point out that the effective range was determined for the first time in~\cite{Li:2024uwu} with the results of Eq.~(\ref{eq:new}). One can see again that the results obtained in~\cite{Li:2024uwu} and those obtained here for the effective range are perfectly compatible within errors. In summary, the results obtained here for $a_{K^+\bar{K}^0}$ provide a precise value for this magnitude, and those for the effective range, determined for the first time in~\cite{Li:2024uwu} and here, offer also a precise determination of this magnitude.

\section{conclusion}
We  investigate the $D^+ \to \bar{K}^0 \pi^+ \eta$ reaction with the purpose of obtaining a precise determination of the  $K^+ \bar{K}^0$  scattering parameters. Our study begins with a comprehensive analysis of the dominant weak decay mechanisms at the quark level, focusing on both external and internal emission contributions. We find that the external emission contributions vanish, providing a clearer understanding of the decay process.
We use the chiral unitary approach, where the $a_0(980)$ resonance is generated by the interaction between the $\pi \eta$ and $K \bar{K}$ channels. This resonance plays a crucial role in shaping the mass distributions, especially in the $\pi^+ \eta$ channel, where we observe a clear peak around 1.0 GeV, corresponding to this resonance.

{Furthermore, our analysis obtained the same important features already discussed in Ref.~\cite{Ikeno:2024fjr}.} These analysis reveals important features of the $\bar{K}^0 \eta$ and $\bar{K}^0 \pi^+$ mass distributions. The $\bar{K}^0 \eta$ distribution exhibits a distinctive double-hump structure due to the interference between the $a_0(980)$ and $K^*_0(1430)$ resonances, with the phase $\phi$ playing a crucial role in determining the shape. On the other hand, the $\bar{K}^0 \pi^+$ distribution is rather structureless.

In addition, we provide an evaluation of the $K^+\bar{K}$ scattering length ($a$) and effective range ($r_0$) using the chiral unitary approach, with values that align well with previous theoretical results.  To estimate the uncertainties in these magnitudes, we used a resampling method, adjusting the parameters of the pion decay constant and the cut-off.
The procedure rendered values of $a$ and $r_0$ and, very importantly, their uncertainties. Together with the results of~\cite{Li:2024uwu} , these values provide the most accurate results at present.

In summary, our work not only provides a detailed understanding of the $D^+ \to \bar{K}^0 \pi^+ \eta$ reaction, but also enhances our knowledge of the underlying dynamics governing the scattering processes in the system. The sharp cusp near the $K^{+} \bar{K}^{0}$ threshold and the double-hump structure in the $\bar{K}^0 \eta$ distribution provide important  insights into the interactions between pseudoscalar meson, and the uncertainties in the parameters are well quantified using the resampling approach.
The method, together with the results of~\cite{Li:2024uwu}, provide the most accurate results for $a$, $r_0$  of $K^{+} \bar{K}^{0}$ at present.

\section{Acknowledgments}
{We would like to sincerely thank Prof. Natsumi Ikeno for the careful revision of this manuscript.}
This work is partly supported by the National Natural Science
Foundation of China under Grants  No. 12405089 and No. 12247108 and
the China Postdoctoral Science Foundation under Grant
No. 2022M720360 and No. 2022M720359. ZY.Yang and J. Song wish to thank support from the China Scholarship Council. This work is also supported by
the Spanish Ministerio de Economia y Competitividad (MINECO) and European FEDER funds under
Contracts No. FIS2017-84038-C2-1-P B, PID2020-
112777GB-I00, and by Generalitat Valenciana under con-
tract PROMETEO/2020/023. This project has received
funding from the European Union Horizon 2020 research
and innovation programme under the program H2020-
INFRAIA-2018-1, grant agreement No. 824093 of the
STRONG-2020 project. This work is supported by the Spanish Ministerio de Ciencia e Innovaci\'on (MICINN) under contracts PID2020-112777GB-I00, PID2023-147458NB-C21 and CEX2023-001292-S; by Generalitat Valenciana under contracts PROMETEO/2020/023 and  CIPROM/2023/59. 
This work is partly supported by the National Natural Science Foundation of China (NSFC) under Grants No. 12365019 and No. 11975083, and by the Central Government Guidance Funds for Local Scientific and Technological Development, China (No. Guike ZY22096024), the Natural Science Foundation of Guangxi province under Grant No. 2023JJA110076.

\bibliography{refs.bib} 
\end{document}